\documentclass[prl,aps,twocolumn,superscriptaddress,notitlepage,10pt]{revtex4-1}

 \pdfoutput=1

\usepackage[english]{babel}
\usepackage[latin1]{inputenc}
\usepackage{hyperref}
\usepackage{amsmath, amssymb, amsfonts,mathrsfs}
\usepackage{amssymb,xcolor}
\usepackage{dsfont}
\usepackage{color}
\usepackage{graphicx}
\usepackage{times}
\usepackage{bbm}
\usepackage{bm}

\DeclareMathOperator{\Id}{\mathds{1}}

\newcommand{\cnot}{\text{CNOT}}

\newcommand{\bra}[1]{\left\langle{#1}\right\vert}
\newcommand{\ket}[1]{\left\vert{#1}\right\rangle}
\newcommand{\proj}[1]{|#1\rangle\langle#1|}

\begin{document}

\title{Experimental generation of entanglement from classical correlations via non-unital local noise}

\author{Adeline Orieux}
\email{adeline.orieux[at]gmail.com} 
\affiliation{Dipartimento di Fisica, Sapienza Universit\`a  di Roma, Piazzale Aldo Moro, 5, I-00185 Roma, Italy}
\affiliation{(present address) T\'el\'ecom ParisTech, CNRS-LTCI, 46 rue Barrault, F-75634 Paris Cedex 13, France}

\author{Mario A. Ciampini}
\affiliation{Dipartimento di Fisica, Sapienza Universit\`a  di Roma, Piazzale Aldo Moro, 5, I-00185 Roma, Italy}

\author{Paolo Mataloni}
\affiliation{Dipartimento di Fisica, Sapienza Universit\`a  di Roma, Piazzale Aldo Moro, 5, I-00185 Roma, Italy}
\affiliation{Istituto Nazionale di Ottica, Consiglio Nazionale delle Ricerche (INO-CNR), Largo Enrico Fermi, 6, I-50125 Firenze, Italy}

\author{Dagmar Bru{\ss}}
\affiliation{Institut f{\"u}r Theoretische Physik III, Heinrich-Heine-Universit{\"a}t D{\"u}sseldorf, D-40225 D{\"u}sseldorf, Germany}

\author{Matteo Rossi}
\affiliation{Dipartimento di Fisica and INFN-Sezione di Pavia, via Bassi 6, I-27100 Pavia, Italy}

\author{Chiara Macchiavello}
\affiliation{Dipartimento di Fisica and INFN-Sezione di Pavia, via Bassi 6, I-27100 Pavia, Italy}

\date{\today}

\begin{abstract}
We experimentally show how classical correlations can be turned into quantum entanglement, via the presence of non-unital local noise and the action of a CNOT gate. We first implement a simple two-qubit protocol in which  entanglement production is not possible  in the absence of local non-unital noise, while entanglement arises with the introduction of noise, and is proportional to the degree of noisiness. We then perform a more elaborate four-qubit experiment, by employing two hyperentangled photons initially carrying only classical correlations. We demonstrate a scheme where the entanglement is generated via local non-unital noise, with the advantage to be robust against local unitaries performed by an adversary.
\end{abstract}

\maketitle

{\it Introduction.} Entanglement is the most precious resource in quantum information processing \cite{nc, dagmar}. However, what is most precious is often also most fragile. Indeed, a profound opponent of entanglement is noise: decoherence and dissipation both typically decrease entanglement, unless they are tailored specifically in a collective way, such as, for example, in correlated noisy channels \cite{correlated,yeo}, or in engineered dissipation of coupled systems \cite{cirac} or in the presence of tunable noise \cite{plenio}. Hence, in general, except for peculiar experimental conditions, noise represents a strong obstacle for entanglement.

\begin{figure*}[ht!] 
\centering
\includegraphics[width=0.9\textwidth]{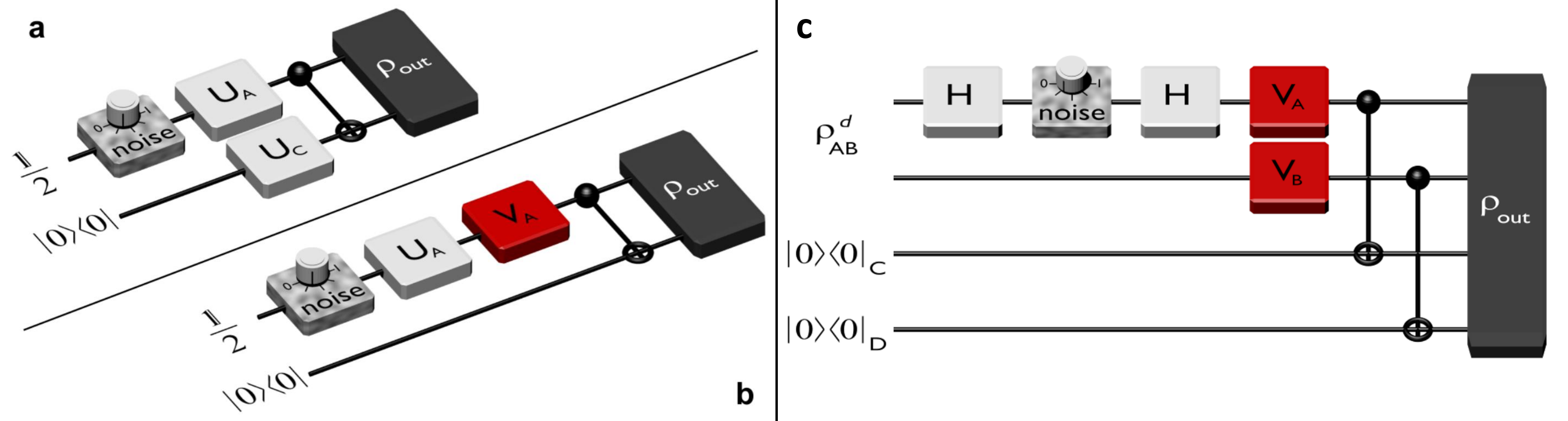}
\caption{a) Two-qubit scheme: the entangled output state $\rho_{out}$ is generated by starting from the initial state $\frac{\mathbb{I}}{2}\otimes |0\rangle \langle 0 |$, and using local unitaries $U_A$ and $U_C$ and a CNOT gate. A local noisy element can be switched on by turning the knob. b) The same protocol in the presence of an adversary player using a local unitary $V_A$ to try and prevent the generation of entanglement.  c) Four-qubit scheme:  here $\rho^d_{AB}$ represents a bipartite state diagonal in the computational basis, with the same block of gates defined in the two-qubit scenario. The aim is to switch on entanglement of $\rho_{out}$ with respect to the cut $AB|CD$ by exploiting local noise.}
\label{fig1}
\end{figure*}

In this work we introduce a scenario where, with fixed given resources of input states and gates, no entanglement can be produced, unless one switches on {\em any} local non-unital noise (such as dissipation). In this setting entanglement is zero without noise, and the degree of the produced entanglement grows with the amount of the introduced local noise. An important aspect of this quantum effect is the choice of the dimension of the underlying Hilbert space: as long as only two qubits are present, we will show that an adversary is always able to prevent the production of entanglement by applying a suitable local rotation.
We will also show that, however, by embedding the protocol into a higher-dimensional Hilbert space, concretely by using four qubits, the creation of  entanglement via local noise can be made robust against any possible unitary action of an adversary. Interestingly, in order to achieve this robustness it is sufficient to have input states that exhibit just classical correlations.
Experimentally, the different dimensional settings of two and four qubits are implemented using the path and polarisation degrees of freedom of photons and the fragility of the effect in low dimensions is demonstrated.

The reason behind the effect described above is an intricate relationship between the concepts of separability, quantum correlations and entanglement: a necessary ingredient for the production of entanglement in our scenario is the generation of nonvanishing off-diagonal terms (``quantum coherences'') in the initial density matrix, which may arise via a local non-unital channel. However, the presence of quantum coherences is not sufficient for robustness of the protocol, in the sense explained above. As mentioned above, a sufficient ingredient for robustness is the presence of classical correlations within the initial state. These classical correlations are turned into correlations of  quantum nature via a local non-unital channel, which still keeps the quantum state separable. Finally, the quantum correlations are activated into entanglement.

Generation of quantum correlations by local noise was theoretically investigated in \cite{alex}, see also \cite{giovanetti,liu}, and experimentally demonstrated with trapped ions \cite{blatt}, while entanglement activation from quantum correlations was theoretically proposed in \cite{piani} (for a different interpretation see \cite{alex2}), and experimentally demonstrated in \cite{acti-exp}. In the scenario that we propose here these two effects are combined, resulting in a scheme where we are able to generate entanglement in a robust way, starting from classically correlated states at the input and switching on just a local noisy device, acting on a single qubit.

{\it Theory.}
We start by briefly describing the setting and the fixed resources for the two-qubit scheme. In the two-qubit protocol (see Fig.~\ref{fig1}a) we are given two qubits in a product state, where a qubit is in the maximally mixed state $\rho_A=\frac{\Id}{2}$ and an ancilla is in the state $\rho_C=\proj{0}$. We are also given a $\cnot$ gate with $A$ ($C$) the control (target) qubit and we are allowed to perform any possible local unitary operation on $A$ and $C$. The goal is to create entanglement in the total state $\rho_{out}$ of qubits $A$ and $C$. It is clear that with only these fixed resources at our disposal it is not possible to create entanglement in the bipartite system considered here.

If in addition we are allowed to use a local noisy device acting only on qubit $A$, such as an amplitude damping channel (see later for details), as reported in Fig.~\ref{fig1}a , the situation changes. Even if at first glance we might expect that our ability to generate entanglement could not be improved by adding such a resource, since the operation is local and noisy channels are usually regarded as detrimental for quantum features, it turns out that if this extra resource is non-unital (i.e. if it does not preserve the identity) it allows to produce an entangled output state $\rho_{out}$.
One can also show that the generation of entanglement in this protocol can always be prevented by a suitable local rotation performed by some adversary before the action of the $\cnot$ gate, as shown in Fig.~\ref{fig1}b. A proof of the above statements is reported in the Supplementary Information (SI).

\begin{figure*}[ht!] 
\centering
\includegraphics[width=0.9\textwidth]{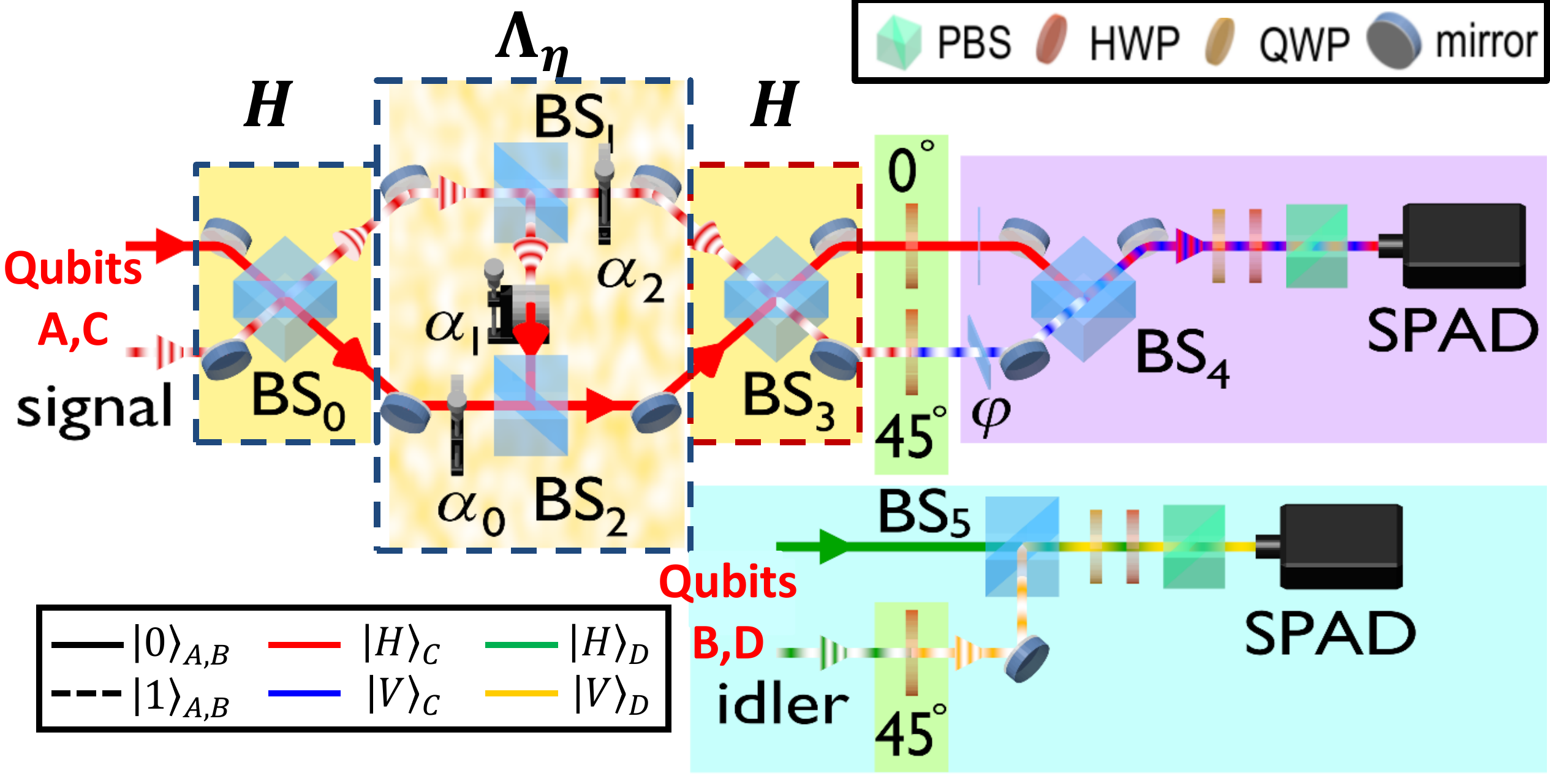}
\caption{Experimental set-up used for 2- and 4-qubit protocol. $BSs$ are balanced beam-splitters, $BS_0$ and $BS_3$ implement the Hadamard gates H; $BS_1$ and $BS_2$, together with the tunable attenuation filters $\alpha_0$, $\alpha_1$ and $\alpha_2$, constitute the amplitude damping channel $\Lambda_\eta$.  Path analysis of photons is performed through the phase plate $\phi$ and beam splitters $BS_4$, $BS_5$. Polarization analysis is performed through standard quantum tomography setup for only the signal photon (2-qubit protocol) and for both signal-idler photons (4-qubit protocol). Here PBS are polarizing beam-splitters, HWP are half-wave plates, QWP are quarter-wave plates. SPAD are single photon avalanche photodiode. The adversary player's attack is simulated by removing $BS_3$ from the set-up.}
\label{fig2}
\end{figure*}

We will now present a 4-qubit scheme where entanglement is generated, starting from a classically correlated state, by again introducing local non-unital noise. Here,  however, the entanglement will not be  vulnerable anymore against local rotations of an adversary.
We begin with a two-qubit state $\rho_{AB}^d$, which is diagonal in the computational basis (which is a factorised basis) and therefore exhibits at most classical correlations, and two ancilla qubits $C$ and $D$, both initialized in the state $\proj{0}$. We are then allowed to apply two $\cnot$ gates that operate on qubits $AC$ and $BD$ (this situation corresponds to the scheme shown in Fig.~\ref{fig1}c when the noisy device is switched off).
Under these conditions the output state in this scenario will always be a separable state \cite{piani}.

We now again introduce a local non-unital noisy device, acting on qubit $A$. The non-unital noise is given by the block of gates $H\Lambda_\eta H$, where $H$ is the Hadamard transform and $\Lambda_\eta$ is an amplitude damping channel, where the noise parameter $\eta$ can be tuned by a knob. The action of the  channel is explicitly described by $\Lambda_\eta(\rho)= \sum_{i=1}^2 A_i^\dagger \rho A_i$, with the Kraus operators $A_1 = \proj{0}+\sqrt{1-\eta}\proj{1}, A_2=\sqrt{\eta}\ket{0}\bra{1}$ with $\eta\in [0,1]$, where $\eta = 0$ corresponds to the noiseless case.
We can show that by adding only this extra local noisy resource entanglement is switched on at the output of the circuit in the bipartition $AB|CD$, starting from states which are at most classically correlated.
We quantify the amount of entanglement in terms of the negativity, defined as $N(\rho) = \sum_i|\lambda_i^-|$, where $\lambda_i^-$ are the negative eigenvalues of the partial transpose of $\rho$ (as mentioned above, we consider here the bipartition $AB|CD$).
It turns out that for any state $\rho_{AB}^d$ the negativity is given by
\begin{equation}
N(\rho_{out})=\frac{\eta}{2},
\label{eq:noise}
\end{equation}
and it  increases by increasing the noise parameter (for more details see SI).

As in the two-qubit  case above, let us introduce an adversary whose goal is to prevent the realization of an entangled output state. In this 4-qubit scheme, we still suppose that she can use only two local unitaries $V_A$ and $V_B$ just before the $\cnot$ gates. Contrary to the simplified two-qubit protocol, the four-qubit protocol turns out to be robust against local unitaries performed by an adversary when the input state is classically correlated (see SI).

{\it Experiment: Two-qubit protocol.}
The 2-qubit protocol was implemented with the optical set-up shown in Fig.~\ref{fig2} where qubits A and C were used.
Qubits $A$ and $C$ were encoded respectively in the path and the polarization degrees of freedom (DOFs) of a single photon generated by a non-linear source of photon pairs \cite{Barbieri}. The input state was prepared with the path qubit in the state $\rho^{d}_A=\frac{1}{2}\proj{0}+\frac{1}{2}\proj{1}$, and the polarization qubit was in the state $\proj{\textrm{H}}_C$, where H (V) designs the horizontal (vertical) polarization and corresponds to the state 0 (1) of the computational basis.
Both Hadamard gates $H$ were realized with balanced beam-splitters (BSs). The amplitude damping channel $\Lambda_\eta$, previously introduced, was achieved by a combination of two balanced BSs and three attenuation filters whose transmission coefficients could be adjusted separately to obtain any value of $\eta \in [0,1]$.
Details on the relationship between $\eta$ and the filters transmission coefficients are given in the SI.
The $\cnot$ gate, controlled by the path qubit with the polarization qubit as target, was implemented by inserting a half-wave plate (HWP) at $45^\circ$ in the path mode corresponding to $\ket{1}_A$. Note that a HWP at $0^\circ$ was also inserted in the other path mode so as to maintain the same optical length for both modes.
Finally, the entanglement in the final state of $AC$ was estimated by a standard tomographic reconstruction measurement of the two-qubit state \cite{PRAkwiat} from which we recovered the negativity $N$. Details on the experiment are given in the SI.

\begin{figure}[h!] 
\centering
\includegraphics[width=88mm]{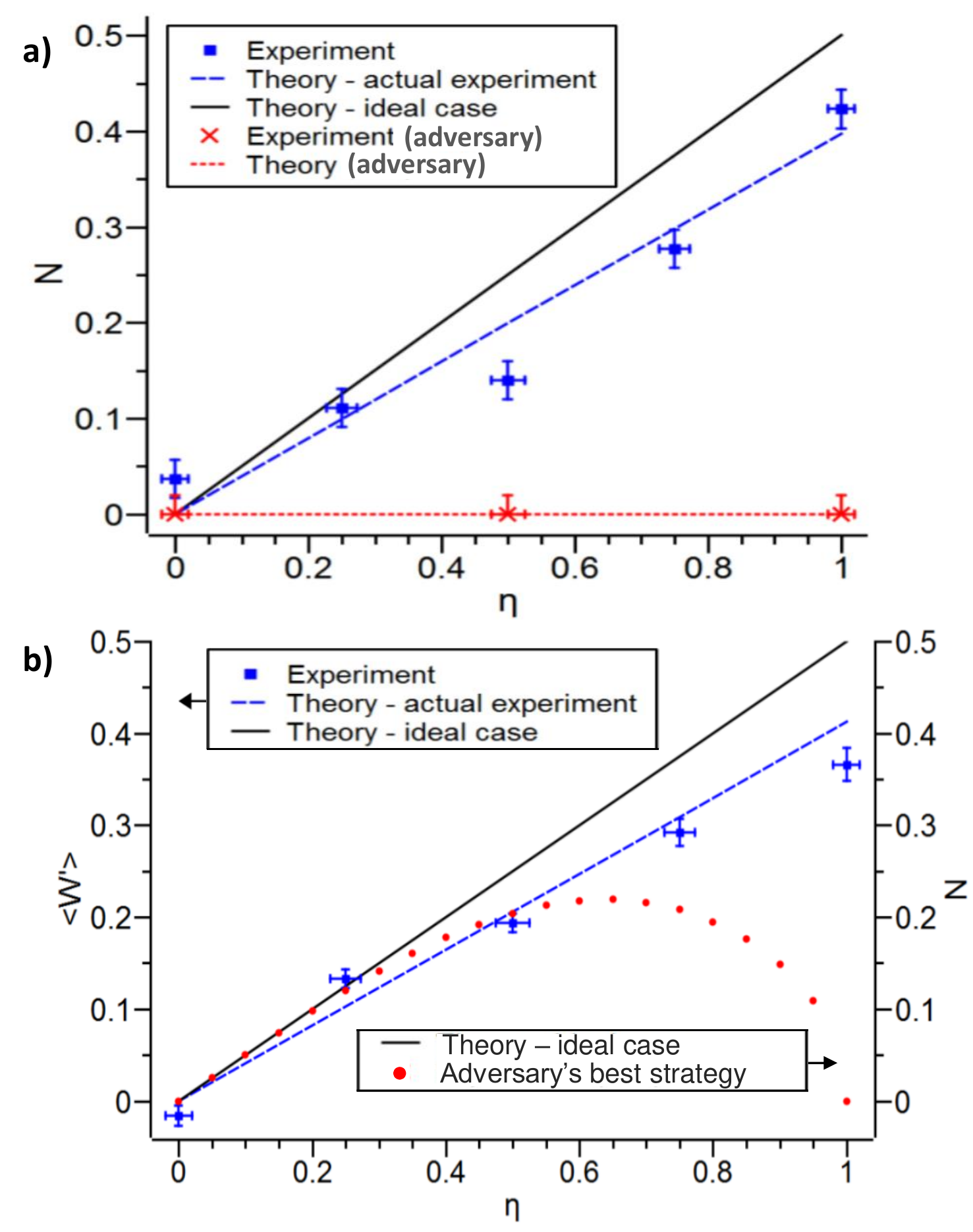}
\caption{Experimental results. a) 2-qubit protocol: negativity $N(\rho_{out})$ vs the noise parameter $\eta$. b) 4-qubit protocol: expectation value $\langle W' \rangle =- \langle W \rangle$ of the entanglement witness applied to $\rho_{out}$ as a function of the noise parameter $\eta$. The dashed lines correspond to a theoretical calculation that takes into account the actual experimental apparatus. The red dots represent the adversary's best strategy. In the 4-qubit case they report a simulation of the minimal amount of entanglement (quantified by the negativity) that can be generated after the application of local rotations on qubits $A$ and $B$ by an adversary. Error bars are calculated from photon counting statistics (see Supplementary Information).}
\label{fig3}
\end{figure}

In Fig.~\ref{fig3}a we report the measured negativity $N$ of the final 2-qubit state obtained for different values of $\eta$ (blue squares), together with its theoretical value given by Eq.~\eqref{eq:noise} (dark line).
The dashed blue line corresponds to the expected theoretical value when we take into account the experimental imperfections of our set-up (see SI).
The figure shows a good agreement between the measured and theoretical behaviors demonstrating that the amount of entanglement created increases as the noise is increased.

As expected, the creation of entanglement can be prevented by Eve for any amount of noise. With the protocol of Fig.~\ref{fig1}b the best strategy for Eve is to perform another Hadamard gate ($V_A=H_{Eve}$) to cancel the action of the second Hadamard. Indeed, in this way, even though the noise is non-unital, the density matrix of qubit $A$ remains diagonal (in the computational basis) and no entanglement is then generated. In order to simulate Eve's attack experimentally we simply removed the beam-splitter $BS_3$ from our set-up (see Fig.~\ref{fig2}). As expected, the negativity measured in this case is always vanishing, as can be seen in Fig.~\ref{fig3}a (red crosses).

{\it Experiment: Four-qubit protocol.}
We implemented experimentally the four-qubit scheme shown in Fig.~\ref{fig2}.
It is based on the previous two-qubit set-up, where we add two extra qubits $B$ and $D$. The four qubits were generated through a source of two photons encoded in the path and the polarization degrees of freedom \cite{Barbieri}. Precisely, qubit $A$ ($B$) is encoded in the path ($k$) DOF of the signal (idler) photon, while qubit $C$ ($D$) is encoded in the polarization ($\pi$) DOF of the signal (idler) photon. The actual experimental setup was built on a chained Sagnac configuration (see SI). The path qubits are prepared in the state $\rho^{d}_{AB}=p\proj{00}+(1-p)\proj{11}$, with $p \simeq 0.5$, and the polarization qubits are both in the state $\proj{\textrm{H}}_{C,D}$.
All the gates were implemented as in the previous two-qubit set-up. The additional $\cnot$ gate between qubits $B$ and $D$ is implemented by inserting a half-wave plate (HWP) at $45^\circ$ in the path mode corresponding to $\ket{1}_B$.

We checked for entanglement in the splitting $AB|CD$ by using the following entanglement witness (see SI):
\begin{align}\label{witness}
W =& \frac{1}{8}\left( \Id \otimes \Id - \sigma_x \otimes \sigma_x + \sigma_y \otimes \sigma_y - \sigma_z \otimes \sigma_z \right)_{AC} \nonumber \\
&\otimes\left( \Id \otimes \Id + \sigma_z \otimes \sigma_z \right)_{BD}
\end{align}
where  $\sigma_x$, $\sigma_y$ and $\sigma_z$ are the Pauli matrices. This witness certifies the presence of bipartite entanglement  in the final state of $ABCD$ whenever $\langle W \rangle < 0$ \cite{detection}.

In Fig.~\ref{fig3}b we show the measured expectation value $\langle W' \rangle =- \langle W \rangle$ of the witness as a function of the noise parameter $\eta$ (blue squares). This quantity provides a lower bound for the negativity of $\rho_{out}$, namely $N(\rho_{out})\ge \langle W' \rangle$ \cite{brandao}.
The blue dashed line corresponds to the theoretical evolution computed by taking into account the experimental imperfections and the actual measured value of $p$ in the input state (see Supplementary Information). Our results are in agreement with Eq.~\eqref{eq:noise}: while there is no entanglement when local noise is absent ($\eta=0$), it is generated as the noise increases.

The experimental demonstration of the robustness of the protocol would require the actual measurement of a non-vanishing entanglement measure for every possible unitary operations that Eve can apply on qubits $A$ and $B$. As this is not practically feasible in our set-up, instead we numerically computed the minimal amount of entanglement that is generated by the protocol when Eve adopts the local rotations on qubits $A$ and $B$ that best reduce this entanglement. This result (see red dots in Fig.~\ref{fig3}b) is obtained, for each value of $\eta$, by a minimisation of the negativity over all possible unitaries $V_A$ and $V_B$.
Notice that when the noise level is low, i.e. $\eta$ is close to zero, Eve's best attack cannot reduce much the amount of created entanglement. Only in the extreme cases $\eta=0, 1$ she can totally prevent the creation of an entangled output state.


{\it Conclusion.}
We have shown that quantum entanglement can be switched on by the help of a local noisy device. The amount of entanglement is also shown to increase by increasing the amount of noise introduced on purpose in the set-up.
Using a single photon, we experimentally demonstrated the creation of entanglement between its path and polarization degrees of freedom demonstrating the idea in terms of a simple two-qubit protocol.
A more sophisticated four-qubit scenario, contrary to the former case, has the advantage of being robust against local unitaries performed by an adversary player. In order to encode all four qubits, we employed two hyperentangled photons, detecting the entanglement of the output state by the use of a suitable witness operator.
Our experiments have thus implemented some subtle features of the quantum world: while, obviously, no local action can produce correlations, a local quantum channel can turn classical correlations into quantum correlations \cite{alex}.
We have realised a local amplitude damping channel that performs this task, while preserving separability of the state. In a second step, the quantum correlations were then  activated into entanglement \cite{piani}, by using non-local gates. We emphasize that without the noisy channel no entanglement could have been produced.
The underlying dimension of the Hilbert space played a crucial role in terms of robustness of the protocol.

\textit{Acknowledgements.} This work was supported by the European Union through the project FP7-ICT-2011-9-600838 (QWAD Quantum Waveguides Application and Development; www.qwad-project.eu).


\end{document}


\title{Supplementary Informations: Experimental generation of entanglement from classical correlations via non-unital local noise}

\author{Adeline Orieux}
\email{adeline.orieux[at]gmail.com} 
\affiliation{Dipartimento di Fisica, Sapienza Universit\`a  di Roma, Piazzale Aldo Moro, 5, I-00185 Roma, Italy}
\affiliation{(present address) T\'el\'ecom ParisTech, CNRS-LTCI, 46 rue Barrault, F-75634 Paris Cedex 13, France}

\author{Mario A. Ciampini}
\affiliation{Dipartimento di Fisica, Sapienza Universit\`a  di Roma, Piazzale Aldo Moro, 5, I-00185 Roma, Italy}

\author{Paolo Mataloni}
\affiliation{Dipartimento di Fisica, Sapienza Universit\`a  di Roma, Piazzale Aldo Moro, 5, I-00185 Roma, Italy}
\affiliation{Istituto Nazionale di Ottica, Consiglio Nazionale delle Ricerche (INO-CNR), Largo Enrico Fermi, 6, I-50125 Firenze, Italy}

\author{Dagmar Bru{\ss}}
\affiliation{Institut f{\"u}r Theoretische Physik III, Heinrich-Heine-Universit{\"a}t D{\"u}sseldorf, D-40225 D{\"u}sseldorf, Germany}

\author{Matteo Rossi}
\affiliation{Dipartimento di Fisica and INFN-Sezione di Pavia, via Bassi 6, I-27100 Pavia, Italy}

\author{Chiara Macchiavello}
\affiliation{Dipartimento di Fisica and INFN-Sezione di Pavia, via Bassi 6, I-27100 Pavia, Italy}

\maketitle

\section{Two-qubit protocol}
In order to understand how the two-qubit protocol works we consider the simplified scenario where we have qubit $A$ in a generic input state $\rho$ (instead of $\frac{\Id}{2}$) and qubit $C$ in state $\ket{0}$, and we act on them with a $\cnot$ gate in the computational basis. This situation is described in Fig.~1a of the main text, where $\rho$ is the resulting state of qubit $A$ after the application of the noisy device and the unitary $U_A$, while $U_C=\Id$.
We then investigate which properties the input state $\rho$ of qubit $A$ has to fulfill in order to obtain an entangled output state $\rho_{out}$.
We quantify the amount of entanglement in the output state by means of the negativity, which is a faithful entanglement measure for a two-qubit system~\cite{vidal}. Within this framework, $\rho_{out}$ is entangled if and only if the input state $\rho$ has non-vanishing off-diagonal terms, i.e. $\rho_{01}\neq 0$. This follows from using the entanglement measure ``negativity'' $N$, defined as
\begin{equation}
N(\rho_{AC}) = \sum_i |\lambda_i^-| ,
\label{negativity}
\end{equation}
where $\lambda_i^-$ are the negative eigenvalues of the matrix resulting from the partial tranposion of $\rho_{AC}$.
It is straightforward to show that the negativity of $\rho_{out}$ is connected to the off diagonal terms of $\rho$ via the formula
\begin{equation}
N(\rho_{out})=|\rho_{01}|.
\label{obs1}
\end{equation}
\noindent Hence, the higher the off-diagonal terms of the input state $\rho$, the more entangled is the bipartite output state $\rho_{out}$. On the other hand, if $\rho$ is diagonal in the computation basis, that is $\rho=p\proj{0}+(1-p)\proj{1}$, the output state turns out to be at most classically correlated, i.e. $\rho_{out}=p\proj{00}+(1-p)\proj{11}$.

The above analysis shows why, within the scheme of Fig.~1a of the main text, we are able to create entanglement by using a non-unital channel while we cannot otherwise: since the state $\frac{\Id}{2}$ is invariant under any unitary transformation $U_A$, it is not possible to produce non-vanishing off-diagonal terms in $\rho_A$ by unitary operations. If, however, a non-unital channel $\Lambda$ is present, it transforms by definition the identity to a different state, i.e. $\Lambda[\frac{\Id}{2}]=\sigma$, where $\sigma\neq \frac{\Id}{2}$. Therefore, by applying a suitable local unitary $U_A$ on $\sigma$ it is possible to generate nonvanishing off-diagonal terms.

Let us now introduce in this scenario an adversary player (Eve) whose aim is to prevent the generation of entanglement in $\rho_{out}$ by performing a unitary operation on qubit $A$ after the state preparation and before the action of the $\cnot$ gate. We can thus imagine that a unitary $V_{A}$ is in Eve's hands, as depicted in  Fig.~1b of the main text. Eve can always win, provided that she knows the quantum state after the action of $U_A$. The reason for this statement is that any single-qubit state can be  made diagonal in the computational basis by applying a suitable rotation.
The two-qubit protocol considered here is therefore not \textit{robust} against local rotations: the generation of entanglement at the end can always be prevented by a suitable local rotation performed by an adversary before the action of the $\cnot$ gate.

\section{Four-qubit protocol}
In order to underline the differences between the two- and four-qubit schemes and in particular to show the robustness of the latter, we refer to a more general scenario where we have a generic input state $\rho_{AB}$ for the first two qubits (thus, not necessarily diagonal in the computational basis) and we analyze what are the properties of the input state that guarantee entanglement at the output state $\rho_{out}$ in the splitting $AB|CD$.
Following the set-up in Fig.~1c  of the main text with a general two-qubit input state $\rho_{AB}$ and no noise (i.e. $\eta=0$), we have that $\rho_{out}$ is entangled iff the input state $\rho_{AB}$ has non-vanishing off-diagonal terms.
This result can be proven by noticing that the negativity of $\rho_{out}$ is connected to the off diagonal terms of $\rho_{AB}$ via the formula \cite{piani}
\begin{equation}
N(\rho_{out})=\sum_{i<j}|\rho_{ij}|.
\label{obs2}
\end{equation}
Moreover, $\rho_{out}$ turns out to be separable whenever $\rho_{AB}$ is diagonal in the computational product basis. Hence, within this set-up, the negativity can be regarded as a faithful entanglement measure. This result explains the role that local noise plays here: it introduces off-diagonal elements in the input state, making the final state $\rho_{out}$ entangled.

We consider now the case where we introduce an adversary (Eve) whose goal is to prevent the realization of an entangled output state by using only two local unitaries $V_A$ and $V_B$ just before the $\cnot$ gates. As shown in Ref. \cite{piani}, Eve's action is always successful iff $\rho_{AB}$ is at most classically correlated. In other words, if we want to create an entangled output state $\rho_{out}$ with certainty in the presence of an adversary  a successful strategy is to prepare an input state $\rho_{AB}$ which has quantum correlations, i.e. correlations that are not strictly classical.

\section{Experimental set-up details}

\subsection{Input state preparation}
The photon pairs used in this work are generated by spontaneous parametric down-conversion (SPDC) in a 1.5~mm-long type-1 $\beta$-barium borate (BBO) crystal, by a CW pump beam at 355~nm. The frequency-degenerate horizontally-polarized SPDC photon pairs are emitted on the surface of a cone, two photons of the same pair being diametrally opposite. To generate the path-qubits, we select with a mask four modes of the emission cone corresponding to two possible pairs: $\ket{00}\bra{00}_{high,low}$ or $\ket{11}\bra{11}_{high,low}$. In this configuration, the source naturally generates the path-entangled Bell state $\ket{\Phi^+}=\frac{1}{\sqrt{2}}(\ket{00}_{high,low}+\ket{11}_{high,low})$ \cite{Barbieri}. In order to ensure that only classical correlations are present in the input state for the experiments at hand, the phase coherence between these two possible pairs is destroyed by having an optical path difference between any two path modes larger than the photons coherence time.

\begin{figure}[h] 
\centering
\includegraphics[width=88mm]{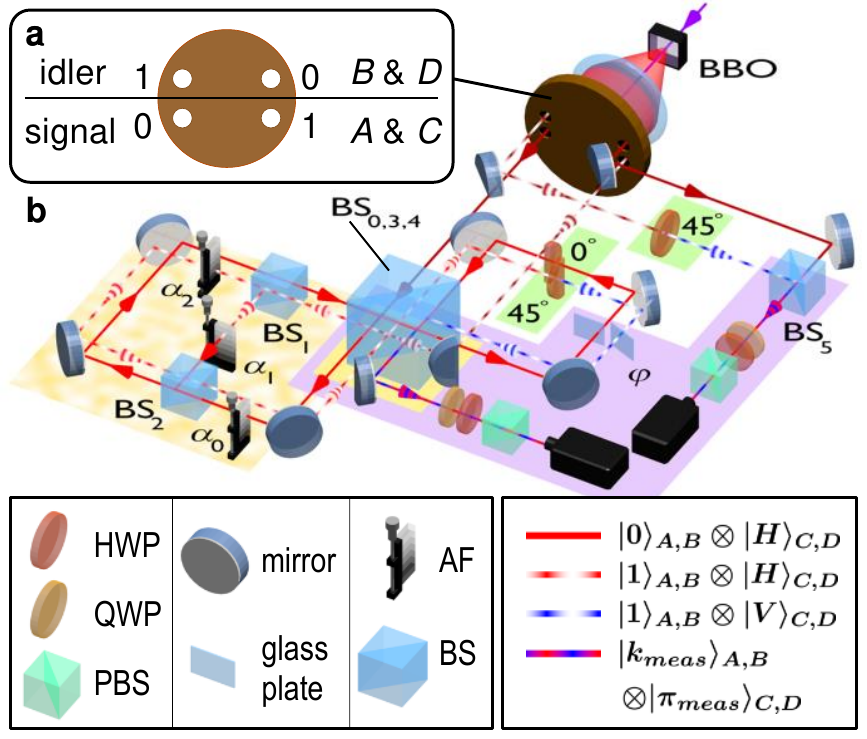}
\caption{{\bf Actual experimental set-up.} {\bf a} Emission modes of the SPDC source. Qubit $A$ ($B$) is encoded in the polarization of the signal (idler) photon, Qubit $C$ ($D$) is encoded in the path of the signal (idler) photon. {\bf b} Actual experimental set-up for the four-qubit protocol (for the two-qubit experiment, the same apparatus is used but the \cnot \, on the idler photon is removed so that this photon serves only as trigger). $\ket{k_{meas}}\otimes\ket{\pi_{meas}}$: path-polarization state of the photon selected by the projective measurement set-up. The central beam-splitter is used three times and plays successively the role of $BS_0$, $BS_3$ and $BS_4$ (see Fig.~2 of the main text).} \label{figMethods} 
\end{figure}
\subsubsection{Amplitude-damping channel}

The experimental implementation of the amplitude-damping channel consists of a combination of two beam-splitters (BS$_1$ and BS$_2$) and three attenuation filters (see the blurred yellow gate in Fig.~2 of the main text, or Fig.~\ref{figMethods}~b) allowing to transfer, in a tunable way, a portion of mode $\ket{1}$ onto mode $\ket{0}$ of the path qubit $A$, encoded in the signal photon. Here we explain the relationship between the noise parameter $\eta$ and the intensity transmission coefficients $\alpha_0$, $\alpha_1$ and $\alpha_2$ of the attenuation filters.

\noindent Let us consider a generic input state (with $\beta,|\gamma|\in[0,1]$)
\begin{equation*}
\rho=(1-\beta)\ket{0}\bra{0}+\beta\ket{1}\bra{1}+\gamma\ket{0}\bra{1}+\gamma^*\ket{1}\bra{0}.
\end{equation*}
The channel $\Lambda_\eta$ (defined by the Kraus operators $\{\proj{0}+\sqrt{1-\eta}\proj{1},\sqrt{\eta}\ket{0}\bra{1}\}$) transforms it into the state
\begin{align*}
\rho'=&(1-(1-\eta)\beta)\ket{0}\bra{0}+(1-\eta)\beta\ket{1}\bra{1}\\
&+\sqrt{1-\eta}\gamma\ket{0}\bra{1}+\sqrt{1-\eta}\gamma^*\ket{1}\bra{0}.
\end{align*}
If we input the same state $\rho$ in the set-up of the blurred yellow gate in Fig.~2 of the main text, we obtain the final state
\begin{align*}
\rho'' =&\xi [ (1-\beta) T \alpha_0 + \beta R^2 \alpha_1 ] \ket{0}\bra{0} + \xi \beta T \alpha_2 \ket{1}\bra{1}\\
&+ \xi \gamma T \sqrt{\alpha_0\alpha_2} \ket{0}\bra{1} + \xi \gamma^* T \sqrt{\alpha_0\alpha_2} \ket{1}\bra{0},
\end{align*}
where $T=57.5\%$ ($R=42.5\%$) is the measured intensity transmission (reflection) coefficient of the beam-splitters and $\xi=((1-\beta) T \alpha_0 + \beta R^2 \alpha_1 + \beta T \alpha_2)^{-1}$ is a normalization factor. Term by term identification between $\rho'$ and $\rho''$ gives:
\begin{eqnarray}
1-(1-\eta)\beta &=& \xi ( (1-\beta) T \alpha_0+\beta R^2 \alpha_1 ),\nonumber\\
(1-\eta)\beta &=& \xi \beta T \alpha_2,\nonumber\\
\sqrt{1-\eta}\gamma &=& \xi \gamma T \sqrt{\alpha_0\alpha_2}. \nonumber
\end{eqnarray}
From these equations, by choosing $\alpha_0=\frac{R^2}{T}=31\%$, we obtain $\alpha_1=\eta$ and $\alpha_2=\alpha_0 (1-\eta)=\frac{R^2}{T}(1-\eta)$. For each setting of $\eta$, the attenuators transmission coefficients are tuned to these values by adjusting their vertical position with respect to the beam (see Fig.~\ref{figMethods}~b).

\subsection{Actual experimental set-up}

In Fig.~\ref{figMethods}~b, we show the experimental implementation of the four-qubit set-up described in Fig.~2 of the main text. It consists in a folded version of the set-up, which both guarantees an intrinsic phase stability for the path qubits and requires less optical elements.

\noindent For the signal photon, the central beam-splitter plays successively the role of both Hadamard gates ($BS_0$ and $BS_3$ in Fig.~2 of the main text) and of the path qubit measurement beam-splitter $BS_4$. Between the Hadamard gates, the photon passes through a first Sagnac loop interferometer, in which the amplitude-damping channel is implemented. Then the \cnot \, gate between qubits $A$ and $C$ is realized by a half-wave plate at 45$^\circ$ on mode $\ket{1}_A$ and a half-wave plate at 0$^\circ$ on mode $\ket{0}_A$. The path qubit is then projected onto the measured state $\ket{k_{meas}}_A$ by the phase plate $\varphi$ and the third passage through the central beam-splitter. Finally, after this second Sagnac loop, the polarization qubit is projected on the measured state $\ket{\pi_{meas}}_C$ by  a quarter- and a half-wave plate and a polarizing beam-splitter. The photon is finally detected on a single-photon avalanche photodiode (SPAD) after spatial filtering with a pinhole and spectral filtering with an interference filter (710~nm central wavelength and 10~nm bandwith).

\noindent On the idler photon, we implement the \cnot \, gate between qubits $B$ and $D$ with a half-wave plate at 45$^\circ$. Both qubits are then projectively measured with the same type of measurement apparatus used for the signal photon.
The same set-up is used for the two-qubit protocol, however in that case the gate \cnot$_{BD}$ is removed and the idler photon plays only the role of a trigger photon.

\section{Robustness of the 4-qubit scheme}

Considering the scheme shown in Fig.~2 of the main text, we show here the robustness of the protocol against local unitaries on $A$ and $B$.
Let us start from a general diagonal state in the computational basis, given by
\begin{equation}\label{input}
\rho_{AB}^d=p\sigma_A^d\otimes\proj{0}+(1-p)\tau_A^d\otimes\proj{1},
\end{equation}
with $\sigma_A^d=q\proj{0}+(1-q)\proj{1}$ and $\tau_A^d=r\proj{0}+(1-r)\proj{1}$. Notice that the above state is at most classically correlated (since it is diagonal in a factorized basis) and  it is factorized (thus it does not contain even classical correlations) iff either $p=0,1$ or $q=r$.
Let us denote the state after the action of the block $H\Lambda_\eta H$ on $A$ by $\rho'_{AB}$, i.e. $\rho'_{AB}=(H\Lambda_\eta H)_A[\rho^d_{AB}]$.
We want first to prove that, whenever the input state $\rho_{AB}^D$ is not factorised (namely when $p\neq 0,1$ and $r\neq q$ in Eq. \eqref{input}), the state $\rho_{AB}'$ is a quantum correlated state, namely its eigenvectors do not provide a factorised bi-orthogonal basis.
By writing the input state $\rho_{AB}^d$ as in Eq.~\eqref{input}, we can rewrite $\rho'_{AB}$ as
\begin{equation}\label{discord}
\rho'_{AB}=p\rho_A'(q)\otimes\proj{0}+(1-p)\rho_A'(r)\otimes\proj{1},
\end{equation}
where $\rho_A'(x)=\frac{1}{2}(\Id+\eta\sigma_x+(2x-1)\sqrt{1-\eta}\sigma_z)$.
After excluding the trivial cases of $p=0,1$, the state (\ref{discord}) is diagonalized in a factorised {bi-orthogonal} basis iff $\rho_A'(q)$ and $\rho_A'(r)$ can be simultaneously  diagonalized. This happens iff they commute, that is iff $[\rho_A'(q),\rho_A'(r)]=0$.
By inserting the general form of $\rho_A'(x)$, we can easily see that the commutativity relation above reduces to
\begin{equation}\label{comm}
2i\eta\sqrt{1-\eta}(q-r)\sigma_y=0,
\end{equation}
which holds only when either $q=r$ or $\eta=0,1$.
By neglecting the noiseless case with $\eta=0$, where $\rho'_{AB}$ is the classically correlated state (\ref{input}), and the case with $\eta=1$, which gives $\rho'_{AB}=\proj{0}\otimes[p\proj{0}+(1-p)\otimes\proj{1}]$, we can see that for initial states $\rho_{AB}$ that are classically correlated ($q\neq r$) the state  $\rho'_{AB}$ is always quantum correlated.
According to Ref. \cite{piani}, the presence of quantum correlations in $\rho'_{AB}$ guarantees that if the state  $\rho'_{AB}$ is the input to the sequence of CNOT gates, then the entanglement production is robust against local unitaries on $A$ and $B$.
Therefore, we can conclude that whenever the input state $\rho_{AB}$ is at least classically correlated the entanglement generation process is guaranteed to be robust.


\section{Entanglement witness in the 4-qubit scheme}

By following the scheme depicted in Fig. \ref{figMethods}a, if we choose as input the state
\begin{equation}
\rho_{in}=\left[p\proj{00}+(1-p)\proj{11}\right]_{AB}\otimes \proj{00}_{CD},
\end{equation}
the output state turns out to be block-diagonal and explicitly given by
\begin{equation}
\rho_{out}=p\rho^+_{AC}\otimes\proj{00}_{BD} +(1-p)\rho^-_{AC}\otimes \proj{11}_{BD}.
\end{equation}
Here we have defined
\begin{equation}
\rho^\pm=
\frac{1}{2}
\begin{pmatrix}
1\pm\sqrt{ \bar \eta} & 0 & 0 & \eta\\
0&0&0&0\\
0&0&0&0\\
\eta & 0 & 0 & 1\mp \sqrt{ \bar \eta}
\end{pmatrix},
\end{equation}
and $\bar \eta=1-\eta$ for convenience. In order to construct a witness operator that detects the entanglement of $\rho_{out}$ with respect to the splitting $AB|CD$, we apply the partial transposition $T_{AB}$ with respect to $AB$, and study the negative eigenvalues of $\rho_{out}^{T_{AB}}$. From simple algebra it turns out that, whenever $\eta\neq 0$, the only two negative eigenvalues are $\lambda_1^-=-p\frac{\eta}{2}$ and $\lambda_2^-=-(1-p)\frac{\eta}{2}$. This proves that $N(\rho_{out})=\frac{\eta}{2}$, independently of $p$. Furthermore, the corresponding eigenvectors are given by
\begin{align*}
\ket{\lambda_1^-}=\ket{\Psi^-}_{AC}\otimes \ket{00}_{BD},\\
\ket{\lambda_2^-}=\ket{\Psi^-}_{AC}\otimes \ket{11}_{BD},\\
\end{align*}
with $\ket{\Psi^-}=\frac{1}{\sqrt 2}(\ket{01}-\ket{10})$.
Therefore, an entanglement witness operator can be defined as
\begin{equation}
W=(\proj{\lambda_1^-}+\proj{\lambda_2^-})^{T_{AB}},
\end{equation}
which, if decomposed in terms of local operators, reads
\begin{align}
W =& \frac{1}{8}\left( \Id \otimes \Id - \sigma_x \otimes \sigma_x + \sigma_y \otimes \sigma_y - \sigma_z \otimes \sigma_z \right)_{AC} \nonumber \\
&\otimes\left( \Id \otimes \Id + \sigma_z \otimes \sigma_z \right)_{BD}
\end{align}
Notice that the experimental detection of $W$ requires only the three measurement settings $\{\sigma_x\sigma_z\sigma_x\sigma_z,\sigma_y\sigma_z\sigma_y\sigma_z,\sigma_z\sigma_z\sigma_z\sigma_z\}$ (in lexicographic order $ABCD$).
